\documentclass{svmult}

\usepackage{makeidx}         
\usepackage{graphicx}        
\usepackage{multicol}        
\usepackage[bottom]{footmisc}
\usepackage{amsmath}
\usepackage{amssymb}
\usepackage{color}

\newcommand{\R}{\mathbb R}

\newcommand{\Lxi}{\mathcal{L}_{\xi}}
\newcommand{\LxiN}{\mathcal{L}_{\xi}^N}

\makeindex             

\begin{document}

\title*{Fixed-Order H-infinity Optimization of Time-Delay Systems}
\author{Suat Gumussoy\inst{1}\and
Wim Michiels\inst{2}}
\institute{Department of Computer Science, K. U. Leuven, Celestijnenlaan 200A, 3001 Heverlee,
        Belgium, \texttt{suat.gumussoy@cs.kuleuven.be}
        \and Department of Computer Science, K. U. Leuven, Celestijnenlaan 200A, 3001 Heverlee,
        Belgium, \texttt{wim.michiels@cs.kuleuven.be}}
%
\maketitle
\vspace{-.1cm}
\begin{abstract}
H-infinity controllers are frequently used in control theory due to their robust performance and stabilization. Classical H-infinity controller synthesis methods for finite dimensional LTI MIMO plants result in high-order controllers for high-order plants whereas low-order controllers are desired in practice. We design fixed-order H-infinity controllers for a class of time-delay systems based on a non-smooth, non-convex optimization method and a recently developed numerical method for H-infinity norm computations.
\end{abstract}
\vspace{-.3cm}
Robust control techniques are effective to achieve stability and performance requirements under model uncertainties and exogenous disturbances \cite{sg_ZhouBook}. In robust control of linear systems, stability and performance criteria are often expressed by H-infinity norms of appropriately defined closed-loop functions including the plant, the controller and weights for uncertainties and disturbances. The optimal H-infinity controller minimizing the H-infinity norm of the closed-loop functions for finite dimensional multi-input-multi-output (MIMO) systems is computed by Riccati and linear matrix inequality (LMI) based methods \cite{sg_DGKF, sg_GahinetApkarianLMI}. The order of the resulting controller is equal to the order of the plant and this is a restrictive condition for high-order plants. In practical implementations, fixed-order controllers are desired since they are cheap and easy to implement in hardware and non-restrictive in sampling rate and bandwidth. The fixed-order optimal H-infinity controller synthesis problem leads to a non-convex optimization problem. For certain closed-loop functions, this problem is converted to an interpolation problem and the interpolation function is computed based on continuation methods \cite{sg_Nagamune}. Recently fixed-order H-infinity controllers are successfully designed for finite dimensional LTI MIMO plants using a non-smooth, non-convex optimization method \cite{sg_GumussoyHIFOO}. This approach allows the user to choose the controller order and tunes the parameters of the controller to minimize the H-infinity norm of the objective function using the norm value and its derivatives with respect to the controller parameters. In our work, we design fixed-order H-infinity controllers for a class of time-delay systems based on a non-smooth, non-convex optimization method and a recently developed H-infinity norm computation method \cite{sg_TW551}. \index{H-infinity optimization} \index{fixed-order controllers}
\vspace{-.5cm}
\section{Problem Formulation} \label{sg_sec:probform}
\vspace{-.2cm}
We consider time-delay plant $G$ determined by equations of the form,
{\small
\begin{eqnarray}
\dot{x}(t)&=&A_0x(t)+\sum_{i=1}^m A_ix(t-\tau_i)+B_1w(t)+B_2u(t-\tau_{m+1}) \label{sg_eq:xeqn}\\
z(t)&=&C_1x(t)+D_{11}w(t)+D_{12}u(t) \label{sg_eq:zeqn} \\
y(t)&=&C_2x(t)+D_{21}w(t)+D_{22}u(t-\tau_{m+2}). \label{sg_eq:yeqn}
\end{eqnarray}}where all system matrices are real with compatible dimensions and \mbox{$A_0\in\R^{n \times n}$}. The input signals are the exogenous disturbances $w$  and the control signals $u$. The output signals are the controlled signals $z$ and the measured signals $y$. All system matrices are real and the time-delays are positive real numbers. In robust control design, many design objectives can be expressed in terms of norms of closed-loop transfer functions between appropriately chosen signals $w$ to $z$.

The controller $K$ has a fixed-structure and its order $n_K$ is chosen by the user \emph{apriori} depending on design requirements,
{\small
\begin{eqnarray}
\dot{x}_K(t)&=&A_Kx_K(t)+B_Ky(t) \label{sg_eq:xkeqn} \\
u(t)&=&C_Kx_K(t) \label{sg_eq:ukeqn}
\end{eqnarray}}where all controller matrices are real with compatible dimensions and $A_K\in\R^{n_K \times n_K}$.

By connecting the plant $G$ and the controller $K$, the equations of the closed-loop system from $w$ to $z$ are written as,\vspace{-.2cm}
{\small
\begin{eqnarray}
\nonumber \dot{x}_{cl}(t)&=&A_{cl,0}x_{cl}(t)+\sum_{i=1}^{m+2} A_{cl,i} x_{cl}(t-\tau_i)+B_{cl} w(t) \label{sg_eq:xcleqn} \\
z(t)&=&C_{cl}x_{cl}(t)+D_{cl}w(t) \label{sg_eq:zcleqn}
\end{eqnarray}}where \vspace{-.3cm}
{\small
\begin{eqnarray}
\nonumber A_{cl,0}&=&\left(
           \begin{array}{cc}
             A_0 & 0 \\
             B_K C_2 & A_K \\
           \end{array}
         \right),\
         A_{cl,i}=\left(
           \begin{array}{cc}
             A_i & 0 \\
             0 & 0 \\
           \end{array}
         \right)\ \rm{for}\ i=1,\ldots,m, \\
\nonumber  A_{cl,m+1}&=&\left(
           \begin{array}{cc}
             0 & B_2 C_K \\
             0 & 0 \\
           \end{array}
         \right), A_{cl,m+2}=\left(
           \begin{array}{cc}
             0 & 0 \\
             0 & B_K D_{22} C_K \\
           \end{array}
         \right), \\
         B_{cl}&=&\left(
           \begin{array}{cc}
             B_1 \\
             B_K D_{21} \\
           \end{array}
         \right),\ C_{cl}=\left(
           \begin{array}{cc}
             C_1 & D_{12} C_K \\
           \end{array}
         \right),\ D_{cl}=D_{11}.
\end{eqnarray}}

The closed-loop matrices contain the controller matrices \mbox{$(A_K, B_K, C_K)$} and these matrices can be tuned to achieve desired closed-loop characteristics.

The transfer function from $w$ to $z$ is,
{\small
\begin{equation} \label{sg_eq:Tzw}
T_{zw}(s) = C_{cl} \left(sI - A_{cl,0} - \sum_{i=1}^{m+2} A_{cl,i}e^{-\tau_i s}\right)^{-1} B_{cl} + D_{cl}
\end{equation}}and we define fixed-order H-infinity optimization problem as the following.

\textbf{Problem}
Given a controller order $n_K$, find the controller matrices $(A_K$, $B_K$, $C_K)$ stabilizing the system and minimizing the H-infinity norm of the transfer function $T_{zw}$.
\vspace{-.5cm}
\section{Optimization Problem}

\subsection{Algorithm}
The optimization algorithm consists of two steps:
\begin{enumerate}
\item {\bf Stabilization:} minimizing the spectral abscissa, the maximum real part of the characteristic roots of the closed-loop system. The optimization process can be stopped when the controller parameters are found that stabilizes $T_{zw}$ and these  parameters are the feasible points for the H-infinity optimization of $T_{zw}$.
\item {\bf H-infinity optimization:} minimizing the H-infinity norm of $T_{zw}$ using the starting points from the stabilization step.
\end{enumerate}

If the first step is successful, then a feasible point for the H-infinity optimization is found, i.e., a point where the closed-loop system is stable. If in the second step the H-infinity norm is reduced in a quasi-continuous way, then the feasible set cannot be left under mild controllability/observability conditions.

Both objective functions, the spectral abscissa and the H-infinity norm, are non-convex and not everywhere differentiable but smooth almost everywhere \cite{sg_Joris}. Therefore we choose a hybrid optimization method to solve a non-smooth and non-convex optimization problem, which has been successfully applied to design fixed-order controllers for the finite dimensional MIMO systems \cite{sg_GumussoyHIFOO}.

The optimization algorithm searches for the local minimizer of the objective function in three steps \cite{sg_BurkeTAC06}:
\begin{enumerate}
\item A quasi-Newton algorithm  (in particular, BFGS) provides a fast way to approximate a local minimizer \cite{sg_LewisBFGS},
\item A local bundle method attempts to verify local optimality for the best point found by BFGS,
\item If this does not succeed, gradient sampling \cite{sg_BurkeHIFOO} attempts to refine the approximation of the local
minimizer, returning a rough optimality measure.
\end{enumerate}

The non-smooth, non-convex optimization method requires the evaluation of the objective function, in the second step this is the H-infinity norm of $T_{zw}$  and the gradient of the objective function with respect to controller parameters where it exists. Recently a predictor-corrector algorithm has been developed to compute the H-infinity norm of time-delay systems \cite{sg_TW551}. We computed the gradients using the derivatives of singular values at frequencies where the H-infinity norm is achieved. Based on the evaluation of the objective function and its gradients, we apply the optimization method to compute fixed-order controllers. The computation of H-infinity norm of time-delay systems (\ref{sg_eq:Tzw}) is discussed in the following section. \index{non-smooth optimization}

\subsection{Computation of the H-infinity Norm}

We implemented a predictor-corrector type method to evaluate H-infinity norm of $T_{zw}$  in two steps (for details we refer to  \cite{sg_TW551}): \index{H-infinity norm computation}
\begin{itemize}
\item {\bf Prediction step:} we calculate the approximate H-infinity norm and corresponding frequencies where the highest peak values in the singular value plot occur.
\item {\bf Correction step:} we correct the approximate results from the predicted step.
\end{itemize}

\subsubsection{Theoretical Foundation}

The following theorem generalizes the well-known relation between the existence of the singular values of the transfer function equal to the fixed value and the existence of the imaginary axis eigenvalues of a corresponding Hamiltonian matrix \cite{sg_Byers} to the time-delay systems:

\begin{theorem} \cite{sg_TW551} \label{sg_thm:GLxi}
Let $\xi> 0$ be such that the matrix
\vspace{-.2cm}
\[
D_{\xi}:=D_{cl}^TD_{cl}-\xi^2 I
\vspace{-.2cm}
\]
is non-singular and define $\tau_{\max}$ as the maximum of the delays $(\tau_1,\ldots,\tau_{m+2})$.
For $\omega\geq 0$, the matrix $T_{zw}(j\omega)$ has a
singular value equal to $\xi>0$ if and only if $\lambda=j\omega$ is an eigenvalue of the linear infinite dimensional operator $\Lxi$ on $X:=\mathcal{C}([-\tau_{\max},\ \tau_{\max}],\mathbb{C}^{2n})$ which is defined by
\vspace{-.2cm}
{\small
\begin{equation}
\mathcal{D}(\mathcal{L}_{\xi}) =\left\{\phi\in X: \phi^{\prime}\in X,\right. \label{sg_def:Lxi2}
\phi^{\prime}(0)=M_{0}\phi(0)  + \sum_{i=1}^{m+2} (M_i\phi(-\tau_i)+M_{-i}\phi(\tau_i) ) \},
\end{equation}}
{\small
\begin{equation}
\mathcal{L}_{\xi}\phi=\phi^{\prime},\;\phi\in\mathcal{D}(\Lxi) \label{sg_def:Lxi}
\end{equation}}
with
{\small
\[
\begin{array}{l}
M_{0}=\left[\begin{array}{cc}  A_{cl,0}-B_{cl} D_{\xi}^{-1} D_{cl}^T C_{cl}& -B_{cl}
D_{\xi}^{-1}B_{cl}^T\\
\xi^2 C_{cl}^T D_{\xi}^{-T} C_{cl} & -A_{cl,0}^T+C_{cl}^TD_{cl} D_{\xi}^{-1} B_{cl}^T
\end{array}\right],\\
M_i=\left[\begin{array}{cc} A_{cl,i} &0\\0&0
\end{array}\right],\ \
M_{-i}=\left[\begin{array}{cc} 0 &0\\0&-A_{cl,i}^T
\end{array}\right],\ \ 1\leq i\leq m+2.
\end{array}
\]}
\end{theorem}

By Theorem \ref{sg_thm:GLxi}, the computation of H-infinity norm of $T_{zw}$ can be formulated as an eigenvalue problem for the linear operator $\Lxi$.
\begin{corollary} \label{sg_cor:Lxi}
${}$ \\
$\|T_{zw}\|_\infty=\sup\{\xi>0: \textrm{operator } \Lxi \textrm{ has an eigenvalue on the imaginary axis}\}$
\end{corollary}

Conceptually Theorem \ref{sg_thm:GLxi} allows the computation of H-infinity norm via the well-known level set method \cite{sg_Boyd, sg_Bruinsma}. However, $\Lxi$ is an infinite dimensional operator. Therefore, we compute the H-infinity norm of the transfer function $T_{zw}$ in two steps:
\begin{enumerate}
  \item[1)] The prediction step is based on a matrix approximation of $\Lxi$.
  \item[2)] The correction step is based on reformulation of the eigenvalue problem of $\Lxi$ as a nonlinear eigenvalue problem of a finite dimension.
\end{enumerate}

The approximation of the linear operator $\Lxi$ and the corresponding standard eigenvalue problem for Corollary~\ref{sg_cor:Lxi} is given in Section \ref{sg_sec:hinfpred}. The correction algorithm of the approximate results in the second step is explained in Section~\ref{sg_sec:hinfcorr}.

\subsection{Prediction Step} \label{sg_sec:hinfpred}
The infinite dimensional operator $\Lxi$ is approximated by a matrix $\LxiN$. Based on the numerical methods for finite dimensional systems \cite{sg_Boyd, sg_Bruinsma}, the H-infinity norm of the transfer function $T_{zw}$ can be computed approximately as
\begin{corollary} \label{sg_LxiN}
${}$ \\
$\|T_{zw}\|_\infty\approx\sup\{\xi>0: \textrm{operator } \LxiN \textrm{ has an eigenvalue on the imaginary axis}\}$.
\end{corollary}

The infinite-dimensional operator $\mathcal{L}_{\xi}$ is approximated by a matrix using a \emph{spectral method} (see, e.g. \cite{sg_Breda}). Given a positive integer $N$, we consider a mesh $\Omega_N$ of $2
N+1$ distinct points in the interval $[-\tau_{\max},\ \tau_{\max}]$:
\begin{equation}\label{sg_defmesh}
{\textstyle \Omega_N=\left\{\theta_{N,i},\ i=-N,\ldots,N\right\}},
\end{equation}
where
\begin{displaymath}
-\tau_{\max}\leq\theta_{N,-N}<\ldots<\theta_{N,0}=0<\cdots<\theta_{N,N}\leq\tau_{\max}.
\end{displaymath}

This allows to replace the continuous space $X$ with the space $X_N$
of discrete functions defined over the mesh $\Omega_N$, i.e.\ any
function $\phi\in X$ is discretized into a block vector
$x=[x_{-N}^T\cdots\ x_{N}^T ]^T\in X_N$ with components
\[{\textstyle
x_i=\phi(\theta_{N,i})\in\mathbb{C}^{2n},\ \ i=-N,\ldots,N}.
\]
Let $\mathcal{P}_N x,\ x\in X_N$ be the unique $\mathbb{C}^{2n}$
valued interpolating polynomial of degree $\leq 2N$ satisfying
\[{\textstyle
\mathcal{P}_N x (\theta_{N,i})=x_{i},\ \ i=-N,\ldots,N}.
\]
In this way, the operator $\mathcal{L}_{\xi}$ over $X$ can be
approximated with the matrix $\mathcal{L}_{\xi}^N:\ X_N\rightarrow
X_N$, defined as
\vspace{-.1cm}
{\small
\begin{eqnarray}
\nonumber \left(\mathcal{L}_{\xi}^N\ x\right)_i&=&\left(\mathcal{P}_N
x\right)^{\prime}(\theta_{N,i}),\quad i=-N,\ldots,-1,1,\ldots,N, \\
\nonumber \left(\mathcal{L}_{\xi}^N\ x\right)_0&=&M_0 \mathcal{P}_N x(0)+\sum_{i=1}^{m+2} (M_i \mathcal{P}_N x(-\tau_i)
+ M_{-i} \mathcal{P}_N x(\tau_i)). \label{sg_defldisc}
\end{eqnarray}}

Using the Lagrange representation of $\mathcal{P}_N x$,
\[{\textstyle
\begin{array}{l}
\mathcal{P}_N x=\sum_{k=-N}^N l_{N,k}\ x_k,\
\end{array}}
\]
where the Lagrange polynomials $l_{N,k}$ are real valued polynomials
of degree $2N$ satisfying
\vspace{-.1cm}
\[{\textstyle
l_{N,k}(\theta_{N,i})=\left\{\begin{array}{ll}1 & i=k,\\
0 & i\neq k,
\end{array}\right.}
\]
 we obtain the explicit form\vspace{-.1cm}
{\small
\[
\mathcal{L}_{\xi}^N=
\left[\begin{array}{lll}
d_{-N,-N} &\hdots & d_{-N,N} \\
\vdots & & \vdots \\
d_{-1,-N} &\hdots & d_{-1,N} \\
a_{-N} & \hdots & a_N\\
d_{1,-N} &\hdots & d_{1,N} \\
\vdots & & \vdots \\
d_{N,-N} &\hdots & d_{N,N}
\end{array}\right]\in\mathbb{R}^{(2N+1)(2n)\times(2N+1)2n},
\]}
where\vspace{-.1cm}
{\small
\[
\begin{array}{lll}
d_{i,k}&=&l^{\prime}_{N,k}(\theta_{N,i}) I,\ \ \ \
i,k\in\{-N,\ldots,N\},\;i\neq0\\
a_0&=& M_0\ x_0+\sum_{k=1}^{m+2} \left(M_k
l_{N,0}(-\tau_k)+M_{-k}l_{N,0}(\tau_k)\right), \\
a_{i}&=&\sum_{k=1}^{m+2} \left(M_k
l_{N,i}(-\tau_k)+M_{-k}l_{N,i}(\tau_k)\right),\ k\in\{-N,\ldots,N\},\ k\neq0.
\end{array}
\]}
\vspace{-.6cm}
\subsection{Correction Step} \label{sg_sec:hinfcorr}

By using the finite dimensional level set methods, the largest level set $\xi$ where $\LxiN$ has imaginary axis eigenvalues and their corresponding frequencies are computed. In the correction step, these approximate results are corrected by using the property that the eigenvalues of the $\Lxi$ appear as solutions of a finite dimensional nonlinear eigenvalue problem. The following theorem establishes the link between the linear infinite dimensional eigenvalue problem for $\Lxi$ and the nonlinear eigenvalue problem.

\begin{theorem} \cite{sg_TW551} \label{sg_thm:Lxi-Hxi}
Let $\xi> 0$ be such that the matrix
\vspace{-.2cm}
\[
D_{\xi}:=D_{cl}^TD_{cl}-\xi^2 I
\vspace{-.2cm}
\]
is non-singular. Then, $\lambda$ is an eigenvalue of linear operator $\Lxi$ if and only if
\vspace{-.2cm}
\begin{equation} \label{sg_prob:HxiEigThm}
\det H_{\xi}(\lambda)=0,
\vspace{-.2cm}
\end{equation}
where
\vspace{-.3cm}
\begin{equation} \label{sg_eq:HamMatrix}
H_{\xi}(\lambda):=\lambda I-M_0-\sum_{i=1}^{m+2} \left(M_i
e^{-\lambda\tau_i}+M_{-i}e^{\lambda\tau_i}\right)
\vspace{-.1cm}
\end{equation}
and the matrices $M_0$, $M_i$, $M_{-i}$ are defined in Theorem \ref{sg_thm:GLxi}.
\end{theorem}

\begin{figure}
\begin{center}
\resizebox{9cm}{!}{\includegraphics{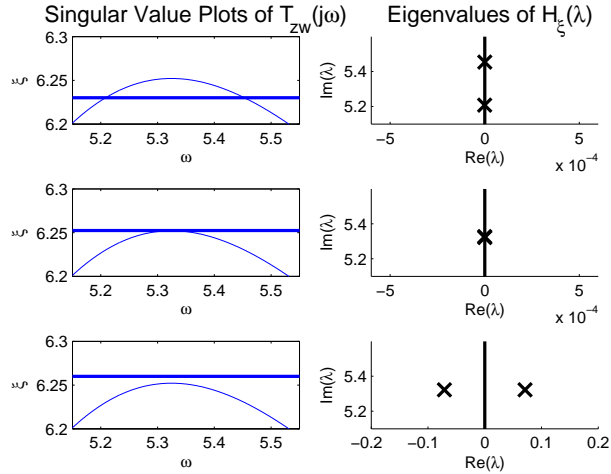}}
 \caption{\label{sg_figD} (left) Intersections of the
 singular value plot of $T_{zw}$ with the horizontal line $\xi=c$,
 for $c<\hat\xi$ (top),\ $c=\hat\xi$ (middle) and
 $c>\hat\xi$ (bottom). (right) Corresponding eigenvalues of $H_\xi(\lambda)$
 (\ref{sg_eq:HamMatrix}).}
\end{center}
\end{figure}

The correction method is based on the property that if \mbox{$\hat{\xi}=\|T_{zw}(j\omega)\|_{\infty}$}, then (\ref{sg_eq:HamMatrix}) has a multiple non-semisimple eigenvalue. If $\hat\xi\geq 0$ and $\hat\omega\geq 0$ are such that
\begin{equation}\label{sg_direct0}
\|T_{zw}(j\omega)\|_{\mathcal{H}_{\infty}}=\hat\xi=\sigma_1(T_{zw}(j\hat\omega)),
\vspace{-.2cm}
\end{equation}
then setting \vspace{-.2cm}
\[
h_{\xi}(\lambda)=\det H_{\xi}(\lambda),
\vspace{-.15cm}
\]
the pair $(\hat\omega,\hat\xi)$ satisfies\vspace{-.1cm}
\begin{equation}\label{sg_direct1}
h_{\xi}(j\omega)=0,\ \ h_{\xi}^{\prime}(j\omega)=0.
\vspace{-.1cm}
\end{equation}
This property is clarified in Figure~\ref{sg_figD}.

The drawback of working directly with (\ref{sg_direct1}) is
that an explicit expression for the determinant of $H_{\xi}$ is
required. This scalar-valued conditions can be equivalently expressed in a matrix-based formulation.
{\small
\begin{equation}\label{sg_forfinal}
{\textstyle \left\{\begin{array}{l}
H(j\omega,\ \xi)\left[\begin{array}{c}u, \\
v\end{array}\right]=0, \quad n(u,v)=0,\\
\Im\left\{v^*\left(I+\sum_{i=1}^{m+1} A_{cl,i}\tau_i e^{-j\omega\tau_i}\right)u\right\}=0\\
\end{array}\right.}
\end{equation}}where $n(u,v)=0$ is a normalizing condition. The approximate H-infinity norm and its corresponding frequencies can be corrected by solving (\ref{sg_forfinal}). For further details, see \cite{sg_TW551}.

\subsection{Computing the Gradients}
The optimization algorithm requires the derivatives of H-infinity norm of the transfer function $T_{zw}$ with respect to the controller matrices whenever it is differentiable. Define the H-infinity norm of the function $T_{zw}$ as
{\small
\[
f(A_{cl,0},\ldots,A_{cl,{m+2}},B_{cl},C_{cl},D_{cl})=\|T_{zw}(j\omega)\|_{\infty}.
\]}These derivatives exist whenever there is a unique
frequency $\hat\omega$ such that (\ref{sg_direct0}) holds,
and, in addition, the largest singular value $\hat\xi$ of
$T_{zw}(j\hat\omega)$ has multiplicity one. Let $w_l$ and
$w_r$ be the corresponding left and right singular
vector, i.e.\
\begin{equation}\label{sg_singvals}
{\small
\begin{array}{l}
T_{zw}(j\hat\omega)\ w_r=\hat \xi\  w_l,\\
w_l^*\ T_{zw}(j\hat\omega)=\hat\xi\ w_r^*.
\end{array}}
\end{equation}
When defining $\frac{\partial f}{\partial A_{cl,0}}$ as a
n-by-n matrix whose $(k,l)$-th element is the derivative
of $f$ with respect to the $(k,l)$-th element of $A_{cl,0}$,
and defining the other derivatives in a similar way, the
following expressions are obtained \cite{sg_Marc}:
{\small
\begin{eqnarray}
\frac{\partial f}{\partial
\nonumber A_{cl,0}}&=&\frac{\Re\left(M(j\hat\omega)^*C_{cl}^Tw_l
w_r^*B_{cl}^TM(j\hat\omega)^{*}\right)}{w_r^*w_r},\\
\nonumber \frac{\partial f}{\partial
A_{cl,i}}&=&\frac{\Re\left(M(j\hat\omega)^*C_{cl}^Tw_l
w_r^*B_{cl}^TM(j\hat\omega)^*e^{j\omega\tau_i}\right)}{w_r^*w_r}\ \textrm{for} \ i=1,\ldots,m+2,\\
\nonumber \frac{\partial f}{\partial B_{cl}}&=&\frac{\Re(
M(j\hat\omega)^*C_{cl}^Tw_lw_r^*)}{w_r^*w_r}, \frac{\partial f}{\partial C_{cl}}=\frac{\Re(w_l w_r^*B_{cl}^T
M(j\hat\omega)^*)}{w_r^*w_r},\\
\frac{\partial f}{\partial
\nonumber D_{cl}}&=&\frac{\Re\left(w_lw_r^{*}\right)}{w_r^*w_r}
\end{eqnarray}}
where $M(j\omega)=\left(j\omega I-A_{cl,0}-\sum_{i=1}^{m+2} A_{cl,i} e^{-j\omega\tau_i}\right)^{-1}$.

We compute the gradients with respect to the controller matrices as
{\small
\begin{eqnarray}
\nonumber \frac{\partial f}{\partial A_K}&=& 
\left[\begin{array}{cc}
0_{n_K \times n} & I_{n_K}
\end{array}\right]
\frac{\partial f}{\partial A_{cl,0}}
\left[\begin{array}{c}
0_{n \times n_K} \\
I_{n_K}
\end{array}\right],\\
\nonumber \frac{\partial f}{\partial B_K}&=& 
\left[\begin{array}{cc}
0_{n_K \times n} & I_{n_K}
\end{array}\right]
\frac{\partial f}{\partial A_{cl,0}}
\left[\begin{array}{c}
I_{n}\\
0_{n_K \times n}
\end{array}\right]C_2^T \\
\nonumber & & \hspace{-.5cm} +
\left[\begin{array}{cc}
0_{n_K \times n} & I_{n_K}
\end{array}\right]
\frac{\partial f}{\partial A_{cl,{m+2}}}
\left[\begin{array}{c}
0_{n \times n_K} \\
I_{n_K}
\end{array}\right]C_K^T D_{22}^T
+\left[\begin{array}{cc}
0_{n_K \times n} & I_{n_K}
\end{array}\right]
\frac{\partial f}{\partial B_{cl}}D_{21}^T ,\\
\nonumber \frac{\partial f}{\partial C_K}&=& 
B_2^T\left[\begin{array}{cc}
I_{n} & 0_{n \times n_K}
\end{array}\right]
\frac{\partial f}{\partial A_{cl,{m+1}}}
\left[\begin{array}{c}
0_{n \times n_K} \\
I_{n_K}
\end{array}\right] \\
\nonumber && +D_{22}^T B_K^T\left[\begin{array}{cc}
0_{n_K \times n} & I_{n_K}
\end{array}\right]
\frac{\partial f}{\partial A_{cl,{m+2}}}
\left[\begin{array}{c}
0_{n \times n_K} \\
I_{n_K}
\end{array}\right]
+D_{12}^T\frac{\partial f}{\partial C_{cl}}
\left[\begin{array}{c}
0_{n \times n_K} \\
I_{n_K}
\end{array}\right]
\end{eqnarray}}
where the matrices $I_n$, $I_{n_K}$  and $0_{n \times n_K}$, $0_{n_K \times n}$ are identity and zero matrices.

\section{Examples}
We consider the time-delay system with the following state-space representation,
{\small
\begin{eqnarray}
\nonumber \dot{x}(t)&=&-x(t)-0.5x(t-1)+w(t)+u(t), \\
\nonumber z(t)&=&x(t)+u(t), \\
\nonumber y(t)&=&x(t)+w(t).
\end{eqnarray}}We designed the first-order controller, $n_K=1$,
{\small
\begin{eqnarray}
\nonumber \dot{x}_K(t)&=&3.61x_K(t)+1.39y(t), \\
\nonumber u(t)&=&-0.83x_K(t)
\end{eqnarray}}achieving the closed-loop H-infinity norm $0.064$. The closed-loop H-infinity norms of fixed-order controllers for $n_K=2$ and $n_K=3$ are $0.021$ and $0.020$ respectively.

Our second example is a $4^{\rm{th}}$-order time-delay system. The system contains $4$ delays and has the following state-space representation,
{\scriptsize
\begin{multline}
\nonumber  \hspace{-.5cm} \dot{x}(t)=
\left(
\begin{array}{cccc}
   -4.4656  &  -0.4271 &   0.4427 &  -0.1854 \\
   -0.8601  & -5.6257  &  0.8577  & -0.5210 \\
    0.9001  & -0.7177  & -6.5358  &  0.0417 \\
   -0.6836  &  0.0242  &  0.4997  & -3.5618
\end{array}
\right)x(t)+ 
\left(
\begin{array}{cccc}
    0.6848 &   -0.0618 &   0.5399 &   0.5057 \\
    0.3259 & -0.3810  &  0.6592  & -0.0066 \\
    0.6325 &   0.3752  &  0.4122  &  0.7303 \\
    0.5878  &  0.9737  &  0.1907 &  -0.8639
\end{array}
\right) x(t-3.2) \\
+
\left(
\begin{array}{cccc}
    0.9371 &  -0.7859  &  0.1332 &   0.7429 \\
   -0.8025 &   0.4483  &  0.6226  &  0.0152 \\
    0.0940  &  0.2274  &  0.1536  &  0.5776 \\
   -0.1941  &  0.5659  &  0.8881 &  -0.0539
\end{array}
\right)x(t-3.4) 
+
\left(
\begin{array}{cccc}
    0.6576 &  -0.8543 &  -0.3460 &   0.6415 \\
   -0.3550 &   0.5024 &   0.6081 &   0.9038 \\
    0.9523  &  0.6624 &   0.0765 &  -0.8475 \\
   -0.4436  &  0.8447 &  -0.0734 &   0.4173 \\
\end{array}
\right)x(t-3.9) \\
+
\left(
\begin{array}{cc}
    1    &     0 \\
   -1.6  & 1 \\
         0   &      0 \\
         0    &     0
\end{array}
\right)w(t)
+
\left(
\begin{array}{cc}
    0.2 \\
   -1 \\
    0.1 \\
   -0.4 \\
\end{array}
\right)u(t-0.2)
\end{multline}}
\vspace{-.3cm}
{\scriptsize
\begin{eqnarray}
\nonumber z(t)&=&
\left(
\begin{array}{cccc}
     1 &    0 &    0 &   -1 \\
     0 &   -1 &    1 &    0
\end{array}
\right)x(t)
+
\left(
\begin{array}{cc}
    0.1 & 1 \\
   -1 &  0.2
\end{array}
\right) w(t)
+
\left(
\begin{array}{c}
    1 \\
   -1
\end{array}
\right) u(t) \\
\nonumber y(t)&=&
\left(
\begin{array}{cccc}
     1 &    0 &    -1 &   0
\end{array}
\right)x(t)
+
\left(
\begin{array}{cc}
    -2 & 0.1
\end{array}
\right) w(t)
+
0.4 u(t-0.2)
\end{eqnarray}}

When $n_K=1$, our method finds the controller achieving the closed-loop H-infinity norm $1.2606$,
{\small
\begin{eqnarray}
\nonumber \dot{x}_K(t)&=&-0.712x_K(t)-0.1639y(t), \\
\nonumber u(t)&=&-0.2858x_K(t)
\end{eqnarray}} and the results for $n_K=2$ and $n_K=3$ are $1.2573$ and $1.2505$ respectively.

\section{Concluding Remarks}

We successfully designed fixed-order H-infinity controllers for a class of time-delay systems. The method is based on  non-smooth, non-convex optimization techniques and allows the user to choose the controller order as desired. Our approach can be extended to general time-delay systems. Although we illustrated our method for a dynamic controller, it can be applied to more general controller structures. The only requirement is that the closed-loop matrices should depend smoothly on the controller parameters. On the contrary, the existing controller design methods optimizing the closed-loop H-infinity norm are based on Lyapunov theory and linear matrix inequalities, which are conservative if the form of the Lyapunov functions are restricted and requires full state information.

\section{Acknowledgements}
This article presents results of the Belgian Programme on Interuniversity Poles of Attraction, initiated by the
Belgian State, Prime Minister's Office for Science, Technology and Culture, the Optimization in Engineering Centre OPTEC of the K.U.Leuven, and the project STRT1-09/33 of the K.U.Leuven Research Foundation.



\printindex
\end{document}